\documentclass[letterpaper,twocolumn,osajnl2,superscriptaddress,twocolumn]{revtex4}
\usepackage{mathptmx}
\usepackage[T1]{fontenc}
\usepackage[latin9]{inputenc}
\usepackage{verbatim}
\usepackage{amsmath}
\usepackage{graphicx}

\makeatletter
\@ifundefined{textcolor}{}
{%
 \definecolor{BLACK}{gray}{0}
 \definecolor{WHITE}{gray}{1}
 \definecolor{RED}{rgb}{1,0,0}
 \definecolor{GREEN}{rgb}{0,1,0}
 \definecolor{BLUE}{rgb}{0,0,1}
 \definecolor{CYAN}{cmyk}{1,0,0,0}
 \definecolor{MAGENTA}{cmyk}{0,1,0,0}
 \definecolor{YELLOW}{cmyk}{0,0,1,0}
 }

\makeatother

\begin{document}

\title{Elliptical dichroism: operating principle of planar chiral metamaterials}

\author{Sergei V. Zhukovsky}

\affiliation{Theoretical Nano-Photonics, Institute of High-Frequency and Communication
Technology, Faculty of Electrical, Information and Media Engineering,
University of Wuppertal, \\
Rainer-Gruenter-Str.~21, D-42119 Wuppertal, Germany, sergei@uni-wuppertal.de.}

\author{Andrey V. Novitsky}

\affiliation{Department of Theoretical Physics, Belarusian State University, Nezavisimosti
Ave. 4, 220030 Minsk, Belarus.}

\author{Vladimir M. Galynsky}

\affiliation{Department of Theoretical Physics, Belarusian State University, Nezavisimosti
Ave. 4, 220030 Minsk, Belarus.}
\begin{abstract}
We employ a homogenization technique based on the Lorentz electronic
theory to show that planar chiral structures (PCSs) can be described
by an effective dielectric tensor similar to that of biaxial elliptically
dichroic crystals. Such a crystal is shown to behave like a PCS insofar
as it exhibits its characteristic optical properties, namely, co-rotating
elliptical polarization eigenstates and asymmetric, direction-dependent
transmission for left/right-handed incident wave polarization.
\end{abstract}


\maketitle
Metamaterials show promise for a wide range of unusual physical phenomena
rare or absent in nature \cite{ShalaevTrans,GiantGyro}, including
giant optical activity \cite{GiantGyro,GiantGyroZhel,CarstenGyro,WegenerGyro,Zhel2003}.
Following a pioneering work by the group of N.~Zheludev in 2006 \cite{ZhelPRL},
planar chiral structures (PCSs) were recently introduced as a distinct
class of metamaterials. They consist of planar elements \cite{ZhelPRL,ZhelPCM-NL1,ZhelPCM-NL2}
that have a sense of {}``twist'' and cannot be superimposed with
their in-plane mirror image. PCSs are asymmetric in electromagnetic
wave propagation for right-handed (RH) vs.~left-handed (LH) circularly
polarized incident wave. Unlike 3D or bilayer chiral metamaterials
\cite{Plum3D,WegenerGyro,CarstenGyro}, which resemble bi-isotropic
or gyrotropic media, PCSs change their properties if wave propagation
direction is reversed \cite{ZhelPRL,ZhelMicroTheory}. They also differ
from Faraday media (also known to have related asymmetry) because
PCSs have \emph{co-rotating} elliptical polarization eigenstates \cite{ZhelPRL,DrezetPCM-OE}
while in bi-isotropic or Faraday media the eigenstates always come
in pairs of RH and LH polarization. 

In this Letter, we interpret optical properties of PCSs in terms of
elliptical dichroism. The properties of chiral metamaterials are usually
investigated on the level of resonant electromagnetic response in
an individual element \cite{Plum3D,ZhelPCM-NL1,ZhelPCM-NL2,ZhelMicroTheory,DrezetPCM-OE,PlumSRRpreprint}.
This heavily depends on the element shape, a wide variety of which
was reported lately \cite{ZhelPRL,ZhelPCM-NL1,ZhelPCM-NL2,DrezetPCM-OE,PlumSRRpreprint}.
Although a recent work \cite{PlumSRRpreprint} offers a generalized
explanation based on polarization-sensitive excitation of electric
and magnetic dipoles, such a treatment should still be performed separately
for different PCS designs. Thus, there is a need for a {}``macroscopic''
description that would make planar chirality available for studies
on an abstract crystallographic level, similar to other optical phenomena
like birefringence or optical activity. 

The idea of regarding elliptical dichroism as a mechanism behind the
optical manifestation of planar chirality is suggested by earlier
experimental accounts of circular dichroism in PCSs \cite{ZhelPRL,PlumSRRpreprint}.
We confirm its existence using the Lorentz-theory homogenization scheme
in chiral split-ring (CSR) structures \cite{PlumSRR,PlumSRR2,PlumSRRpreprint}.
Moreover, we show that a bulk elliptically dichroic medium exhibits
all the characteristic properties of PCS, namely (i) co-rotating elliptical
polarization eigenstates, (ii) asymmetric transmission for RH vs.
LH circular polarization, and (iii) enantiomeric asymmetry and change
of properties for different wave propagation direction. %
{}

\begin{figure}[b]
\includegraphics[width=0.77\columnwidth]{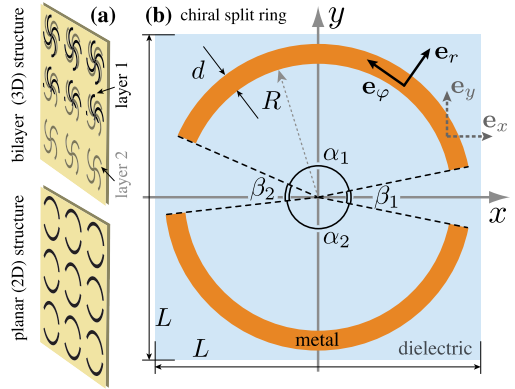}

\caption{(a) Bilayer (3D) vs.~monolayer (planar) chiral metamaterial. (b)
Unit cell of a CSR structure \cite{PlumSRRpreprint}. \label{fig:CSR}}

\end{figure}

\begin{figure}
\includegraphics[width=1\columnwidth]{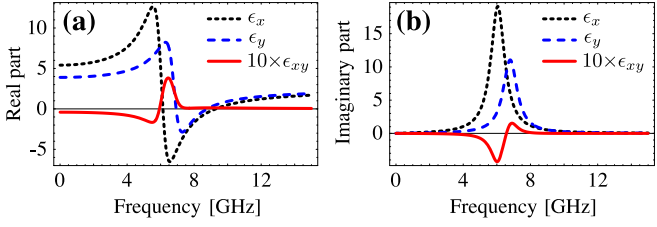}

\caption{Frequency dependencies for (a) real and (b) imaginary parts of~$\hat{\varepsilon}_{\text{eff}}$
in Eq.~\eqref{eq:eps_matrix} for $L=15\text{ mm}$, $R=6\text{ mm}$,
$d=0.8\text{ mm}$, $\alpha_{1}=140^{\circ}$, $\alpha_{2}=160^{\circ}$,
$\beta_{1}=40^{\circ}$, $\beta_{2}=20^{\circ}$. \label{fig:components}}

\end{figure}

We begin by considering a CSR unit cell (Fig.~\ref{fig:CSR}b). For
planar (as opposed to 3D) elements, the electric response is dominant
because the magnetic moment is parallel to the wave propagation direction
\cite{ZhelPRL}. Hence, one can arrive at homogenized material properties
by considering the response within the framework of the electronic
Lorentz theory by using $\hat{\varepsilon}_{\textrm{eff}}\mathbf{E}=\mathbf{E}-\left(4\pi/\mathrm{i}\omega\right)e\left\langle N\mathbf{v}\right\rangle $.\textbf{
}$N$~and~$\mathbf{v}$ are the electron concentration and velocity
\cite{JacksonBook}, which are position-dependent in a structured
material. To derive the effective dielectric tensor~$\hat{\varepsilon}_{\textrm{eff}}$,
one needs to average over the unit cell. If $d\ll R$, the electrons
in the ring can only be mobile in the azimuthal direction so that
$\mathbf{v}_{m}=R\dot{\varphi}\mathbf{e}_{\varphi}$ and $\ddot{\varphi}+\gamma\dot{\varphi}+\omega_{0}^{2}\varphi=\left(e/Rm_{e}\right)\mathbf{e}_{\varphi}\cdot\mathbf{E}_{m}$.
The electrons in the dielectric substrate are bound and their velocity~$\mathbf{v}_{d}$
in response to the electric field is position independent. The averaging
is then performed as\begin{equation}
\left\langle N\mathbf{v}\right\rangle =L^{-2}\left(\int_{\text{metal}}N_{m}\mathbf{v}_{m}(\varphi)\mathrm{d}^{2}\mathbf{r}+\int_{\text{diel}}N_{d}\mathbf{v}_{d}\mathrm{d}^{2}\mathbf{r}\right).\label{eq:homogenization}\end{equation}
For the structure shown in Fig.~\ref{fig:CSR}, integration in Eq.~\eqref{eq:homogenization}
results in a complex symmetric tensor~$\hat{\varepsilon}_{\textrm{eff}}$\begin{equation}
\hat{\varepsilon}_{\textrm{eff}}(\omega)=\left[\begin{array}{ccc}
\epsilon_{x} & \epsilon_{xy} & 0\\
\epsilon_{xy} & \epsilon_{y} & 0\\
0 & 0 & \epsilon_{z}\end{array}\right].\label{eq:eps_matrix}\end{equation}
Its components depend on dielectric constants of metal $\epsilon_{m}(\omega)=1+\omega_{p}^{2}/\left(\omega_{0}^{2}-\omega^{2}-\mathrm{i}\omega\gamma\right)$
and dielectric $\epsilon_{d}(\omega)=1+\Omega_{p}^{2}/\left(\Omega_{0}^{2}-\omega^{2}-\mathrm{i}\omega\Gamma\right)$,
as well as on the geometrical parameters, e.g., the angles $\alpha_{1,2}$~and~$\beta_{1,2}$
(Fig.~\ref{fig:CSR}). Fig.~\ref{fig:components} shows the dependencies
$\epsilon_{x,y,xy}(\omega)$. If the split ring is symmetric ($\beta_{2}=0$,
$\alpha_{1}=\alpha_{2}$, or $\beta_{1}=\beta_{2}$), then $\epsilon_{xy}=0$
and the tensor~\eqref{eq:eps_matrix} supports two linearly polarized
eigenwaves different in their attenuation (linear dichroism). Such
CSRs display no planar chirality \cite{PlumSRR,PlumSRR2}. Otherwise
(when planar chirality is present), $\epsilon_{xy}\neq0$ and the
eigenpolarizations are elliptical, (the effective medium is elliptically
dichroic). %
{}We further rewrite the tensor~$\hat{\varepsilon}_{\textrm{eff}}$
in axial representation (assuming a monoclinic crystal as a particular
case of biaxial media \cite{FedorovBook})\textbf{ }as\begin{equation}
\hat{\varepsilon}=\epsilon_{o}+\left(\epsilon_{e}-\epsilon_{o}\right)\mathbf{c}\otimes\mathbf{c},\quad\mathbf{c}=\mathbf{c}'+\mathrm{i}\kappa\mathbf{c}'',\label{eq:eps_ansatz}\end{equation}
where the vectors $\mathbf{c}'$~and~$\mathbf{c}''$ determine the
optical axes, and $\mathbf{c}\otimes\mathbf{c}$ denotes the outer
dyadic product $\left(\mathbf{c}\otimes\mathbf{c}\right)_{ij}\equiv c_{i}c_{j}$. 

At any fixed frequency, the medium's polarization eigenstates can
be recovered from its normal refraction tensor~$\mathcal{N}_{H}$
\cite{BorzdovTheory}. For non-magnetic media and normal incidence\begin{equation}
\mathcal{N}_{H}^{2}=-\mathbf{q}^{\times}\hat{\varepsilon}\mathbf{q}^{\times}+\left(\mathbf{q}\hat{\varepsilon}\mathbf{q}\right)^{-1}\mathbf{q}^{\times}\hat{\varepsilon}\mathbf{q}\otimes\mathbf{q}\hat{\varepsilon}\mathbf{q}^{\times},\label{eq:NH}\end{equation}
where $\mathbf{q}$ is a unit vector normal to the plane and $\mathbf{q}^{\times}$
is defined as $\left(\mathbf{q}^{\times}\right)\mathbf{u}\equiv\mathbf{q}\times\mathbf{u}$.
The eigenvectors $\mathbf{H}_{1,2}$ of $\mathcal{N}_{H}^{2}$ represent
the polarization states of the field that are preserved as the wave
propagates along~$\mathbf{q}$. If the polarization is circular or
elliptical, the %
{}handedness of~$\mathbf{H}_{j}$ is determined by the sign of the
product $\mathbf{q}\cdot\left[\textrm{Re}\,\mathbf{H}_{j}\times\textrm{Im}\,\mathbf{H}_{j}\right]$
\cite{FedorovBook}. The condition for co-rotation of polarization
eigenstates then becomes\begin{equation}
\eta=\left(\mathbf{q}\cdot\left[\textrm{Re}\,\mathbf{H}_{1}\times\textrm{Im}\,\mathbf{H}_{1}\right]\right)\left(\mathbf{q}\cdot\left[\textrm{Re}\,\mathbf{H}_{2}\times\textrm{Im}\,\mathbf{H}_{2}\right]\right)>0.\label{eq:criterion}\end{equation}

For anisotropic media without absorption, $\eta=0$ (the eigenstates
have linear polarization). For bi-isotropic chiral or Faraday media,
it can be shown that $\eta<0$, so circular or elliptical polarization
eigenstates are counter-rotating. However, for~$\hat{\varepsilon}$
of Eq.~\eqref{eq:eps_ansatz} where $\mathbf{c}'\equiv\mathbf{e}_{x}\cos\phi+\mathbf{e}_{y}\sin\phi$
and $\mathbf{c}''\equiv\mathbf{e}_{x}\cos\psi+\mathbf{e}_{y}\sin\psi$,
Eqs.~\eqref{eq:NH} and~\eqref{eq:criterion} result in\begin{equation}
\eta=\frac{4\kappa^{2}\sin^{2}(\phi-\psi)}{(1+\kappa^{2})^{2}-(\cos2\phi+\kappa^{2}\cos2\psi)^{2}}\geq0.\label{eq:eta}\end{equation}

\begin{figure}
\includegraphics[width=0.8\columnwidth]{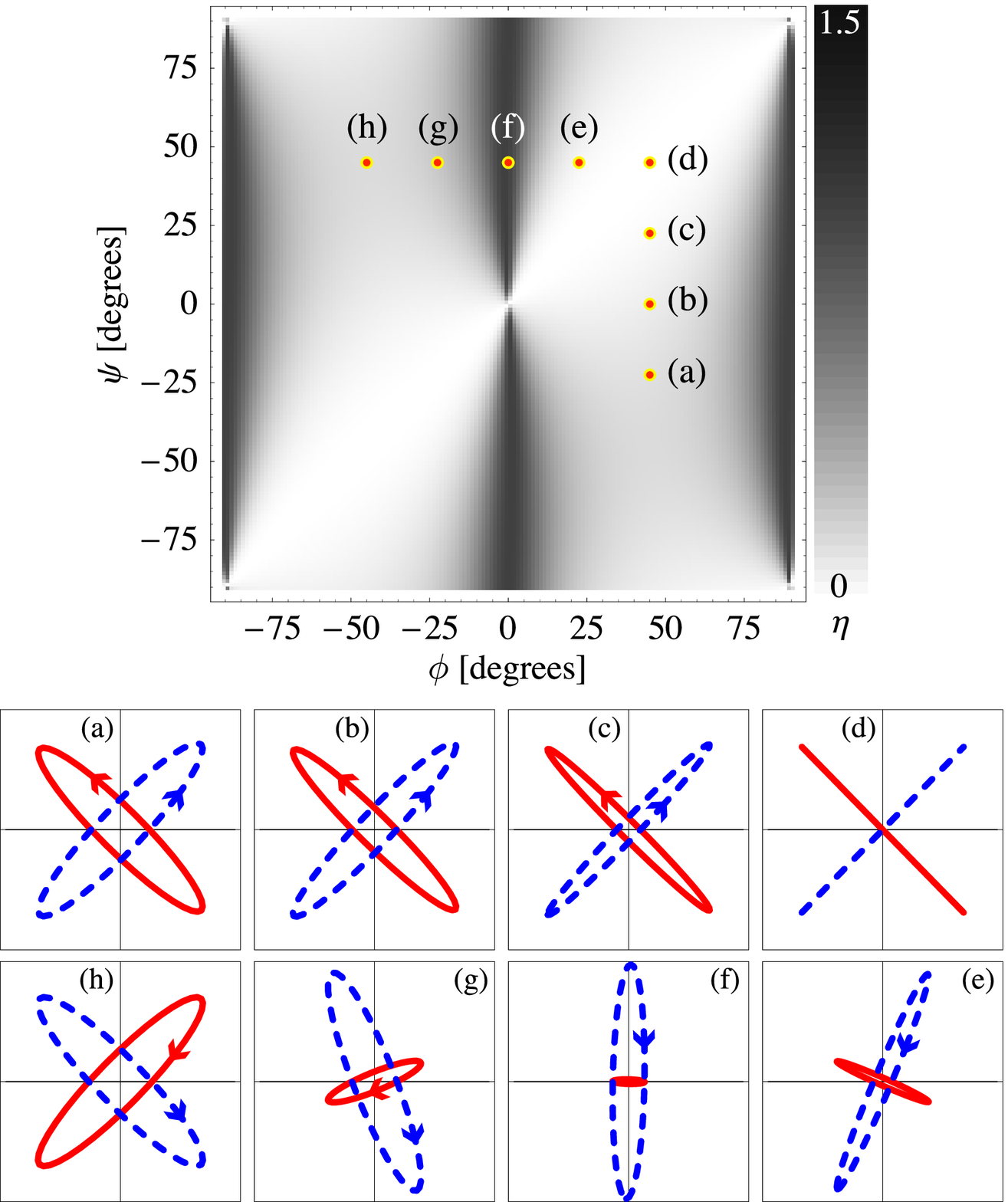}

\caption{The dependence $\eta(\phi,\psi)$ as in Eq.~\eqref{eq:eta} for $\kappa=0.2$
with polarization eigenstates given by Eq.~\eqref{eq:NH} at points
(a)--(h). \label{fig:eigenstates} }

\end{figure}

Fig.~\ref{fig:eigenstates} shows the dependence of~$\eta$ on the
orientation of the axes $\mathbf{c}'$~and~$\mathbf{c''}$. One
can see that polarization eigenstates are indeed co-rotating. They
reverse their handedness at the line $\phi=\psi$%
{}. Note that including gyrotropy in~$\hat{\varepsilon}$ %
{}would result in a negative addition to~$\eta$, thus counteracting
planar chirality as pointed out earlier \cite{DrezetPCM-OE}. 

\begin{figure}
\includegraphics[width=1\columnwidth]{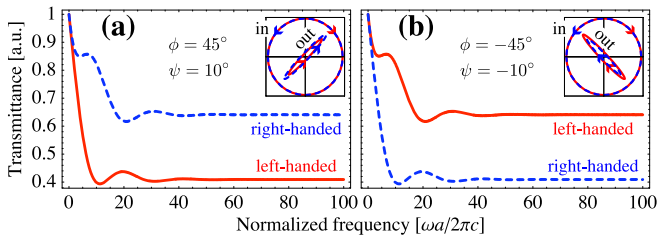}

\caption{Transmission spectra for an LH/RH circularly polarized wave passing
through a slab with thickness~$a$ made of material as in Eq.~\eqref{eq:eps_ansatz}
for $\phi=\pm45^{\circ},\psi=\pm10^{\circ}$, $\epsilon_{o}=1.1$,
$\epsilon_{e}=1.0$, $\kappa=0.2$, along with the incident (in) and
transmitted (out) wave polarization at $\omega a/2\pi c=100$. \label{fig:asymmetry}}

\end{figure}

Next, we consider the transmission spectra of an elliptically dichroic
layer, which can be obtained by constructing the wave evolution operator
in a generalized transfer matrix formalism \cite{BorzdovTheory}.
The Lorentzian resonance is assumed to be broad enough so that material
dispersion can be neglected. Fig.~\ref{fig:asymmetry} shows the
spectra for the LH/RH circularly polarized incident wave. The transmission
of LH vs.~RH polarization is clearly asymmetric. For higher frequencies,
the spectra become flat. This is caused by the absence of material
dispersion, so that spectrally uniform elliptical dichroism results
in spectrally uniform planar chiral behavior. Real metamaterials are
dispersive and thus elliptically dichroic only in a narrow spectral
range close to resonance (see Fig.~\ref{fig:components}). Hence
real PCSs display planar chirality in a resonant manner, in accordance
with earlier findings \cite{ZhelPRL,PlumSRRpreprint}. 

If the material is replaced with its enantiomeric counterpart ($\phi\to-\phi$,
$\psi\to-\psi$), or if the wave propagates in the opposite direction
($\mathbf{q}\to-\mathbf{q}$), the LH/RH plots are exchanged (Fig.~\ref{fig:asymmetry}).
This agrees with experimental results for PCSs \cite{PlumSRRpreprint}
and with the Lorentz homogenization scheme. 

When an initially circularly polarized wave travels through the layer,
the wave is first seen to experience polarization mixing similar to
what happens in a birefringent crystal (LH~$\leftrightarrow$~RH).
This is associated with the oscillatory portion in the spectra (Fig.~\ref{fig:asymmetry}).
 As the wave propagates further, it assumes the polarization matching
one of the eigenstates in orientation and both eigenstates in handedness.
So, a wave whose handedness does not match that of the eigenstates
undergoes polarization conversion, as reported earlier \cite{ZhelPRL}.
This can be understood from the combined action of anisotropy (which
creates polarization mixing) and dichroism (which diminishes the circular
polarization component not matching the eigenstates in handedness).%
{}

The results obtained are complementary to a recent account \cite{PlumSRRpreprint}
linking planar chirality with polarization-sensitive excitation of
resonant modes in CSRs. There, e.g., LH and RH waves were shown to
excite an electric dipole (well coupled to the field) and a magnetic
dipole (poorly coupled), respectively. For the latter, the energy
is {}``trapped'' in the resonant mode and dissipates, resulting
in circular dichroism. The range of PCS designs where dichroism is
brought about by this mechanism remains to be determined. It is also
interesting to note that in high-$T_{c}$ {}``anyon superconductors''
circular dichroism was reported \cite{AnyonLyons} yet no sign of
Faraday-like non-reciprocity was found \cite{AnyonSpielman}, similar
to PCSs. This striking similarity may have fundamental physical reasons
given that such superconductors are known to possess layered geometrical
structure.

To summarize, we have shown that bulk media with elliptical dichroism
exhibit optical properties characteristic for PCSs. First, the elliptical
polarization eigenstates are co-rotating (Fig.~\ref{fig:eigenstates}).
Second, transmission for LH/RH-polarized incident wave is asymmetric
and is exchanged if the material is replaced with its enantiomeric
counterpart (Fig.~\ref{fig:asymmetry}). Finally, an elliptically
polarized wave propagating in such a material undergoes circular polarization
conversion so as to match the polarization eigenstates. The medium
is described solely by a complex symmetric dielectric tensor whose
structure is given by homogenization of a split-ring PCS. This way,
it is shown that a crystallographic approach is possible in studying
PCSs, offering a theoretical starting point for analyzing planar chirality
as an optical phenomenon. One can use existing bianisotropic multilayer
solvers \cite{ourJOA} to efficiently model PCS-based optical devices,
with possible applications in polarization-sensitive integrated optics. 

The authors acknowledge helpful suggestions from V.~Fedotov and E.~Plum
and are grateful to D.~N.~Chigrin for critical reading of the manuscript.
This work was supported in part by the Deutsche Fouschungsgemeinschaft
(FOR 557) and the Basic Research Foundation of Belarus (F08MS-006).

\end{document}